\documentclass{moriond}

\usepackage{amsmath}
\usepackage{amssymb}

\def\be{\begin{equation}}
\def\ee{\end{equation}}
\def\bea{\begin{eqnarray}}
\def\eea{\end{eqnarray}}

\newcommand{\Photo}{}

\begin{document}
\vspace*{4cm}
\title{Two Higgs doublet models with local $U(1)_H$ gauge symmetry and dark matter}

\author{P. Ko}

\address{School of Physics, KIAS, Seoul 130-722, Korea}

\author{Yuji Omura}

\address{Department of Physics, Nagoya University, Nagoya 464-8602, Japan}

\author{Chaehyun Yu}

\address{School of Physics, KIAS, Seoul 130-722, Korea}

%\maketitle\abstracts{
\maketitle
\abstracts{
%The Higgs-meidated flavor changing neutral current (FCNC) problem in two Higgs 
%doublet models (2HDM) is usually resolved by imposing softly broken $Z_2$ discrete %symmetry.   
We propose to implement the softly broken  $Z_2$ symmetry in the usual  two Higgs 
doublet model (2HDM) to spontaneously broken local $U(1)_H$ gauge symmetry, 
and show that the resulting phenomenology can be very rich and is distinctly different from 
the usual 2HDMs. % with $Z_2$ symmetry.  
Likewise, the exact $Z_2$ symmetry in ordinary inert doublet model (IDM) could be 
a remnant of spontaneously broken local $U(1)_H$ symmetry stabilizing 
%, and 
%would be the origin of $Z_2$ symmetry, guarantees the stability of 
dark matter (DM).  In this case, new channels for DM pair 
annihilation into $U(1)_H$ gauge boson(s) open up, allowing the DM mass below 
$\lesssim 40$ GeV, unlike the usual IDM.  
}

\section{Two Higgs doublet Models}
After the discovery of a new boson at the LHC~\cite{higgsdiscovery},
the most important task in particle physics phenomenology would be
the precise measurements of properties of the new boson.
Up to now, the boson looks like a Standard-Model(SM)-like Higgs boson~\cite{Chpoi:2013wga}.
However, this SM-like Higgs boson could be one of Higgs bosons from the
extended Higgs sectors rather than the SM Higgs boson. 
%in model where the Higgs sector in the SM is extended.
In fact, a lot of high energy theories like SUSY, GUT {\it etc}  
predict various extensions of the SM Higgs doublet, such as 
two Higgs doublet models (2HDMs).
%In this respect, 
%Among them, two Higgs doublet models (2HDMs) could be
%effective models of high-energy theories.

%On the other hand, 
The two Higgs double model itself is quite interesting.
It has a lot of scalar bosons, which have rich phenomenology
at the LHC.  Also depending on the setup,  the models could have interesting 
connections to  dark matter physics, baryon number asymmetry of the universe, and
neutrino mass generation. The recent experimental anomalies
in the top forward-backward asymmetry at the Tevatron
and in $B\to D^{(\ast)}\tau\nu$ branching ratios at BABAR
might be explained in non minimal flavor violating 2HDM with flavor dependent
chiral $U(1)$ interactions~\cite{chiralu1}.

In generic 2HDMs, there appear Higgs-mediated  flavor changing neutral current 
(FCNC) problems if both doublets couple to the SM fermions. 
These problems are typically avoided by assigning ad hoc $Z_2$ discrete 
symmetry. With proper $Z_2$ parity assignments to the SM fermions
and two Higgs doublets, one can construct 2HDMs without FCNCs at the tree level.
However, discrete symmetry could generate a domain wall problem when
it is spontaneously broken. Commonly, softly broken $Z_2$ terms are added 
to the models, but the origin of this soft breaking of $Z_2$ is rather unclear.
Recently, it was proposed that the softly broken $Z_2$ symmetry can be replaced by
local $U(1)_H$ gauge symmetry associated with Higgs flavors~\cite{2hdm}.
The local $U(1)_H$ gauge symmetry could be the origin of 
$Z_2$ symmetry in the usual 2HDM, and the softly broken terms are generated
after $U(1)_H$ symmetry breaking. One example is given by the $U(1)_H$-invariant
term, $H_1^\dagger H_2 \Phi$, where $\Phi$ is SM-singlet and $U(1)_H$-charged.
After symmetry breaking, $\Phi$ has a vacuum expectation value (VEV)
and the $H_1^\dagger H_2$ term is generated.

By proper $U(1)_H$ charge assignments  to the SM fermions, one can construct 
new 2HDMs with $U(1)_H$ symmetry (denoted by 2HDMw$U(1)_H$), 
which get reduced to the usual Type-I, -II, -X, and -Y 2HDMs~\cite{2hdm} 
when the $U(1)_H$ gauge boson becomes very heavy. 
In the Type-I case, it is possible to construct the 2HDMw$U(1)_H$ without extra chiral 
fermions except right-handed neutrinos.
For example, assigning zero charges to all the SM fermions in Type-I 2HDM, 
the model becomes anomaly-free 2HDM. In this case, the new gauge boson $Z_H$ 
does not couple  to the SM fermions and it becomes naturally fermiophobic.
However, for Type-II or other cases, extra chiral fermions are required to cancel
gauge anomalies involving the new $U(1)_H$.
And one of the extra chiral fermions could be a good candidate for cold dark 
matter (CDM).

%--------------------------------------------------------------------------
\begin{figure}[th]
%\begin{minipage}{0.5\linewidth}
\centerline{\includegraphics[width=0.40\linewidth]{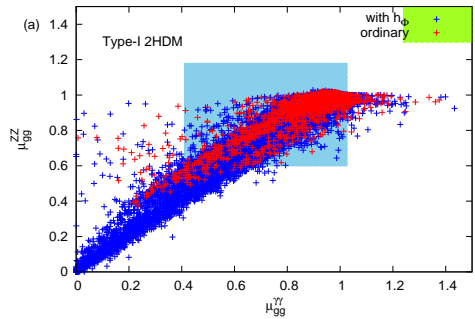}}
%\end{minipage}
%\hfill
\caption[]{
Signal strengths $\mu_{gg}^{\gamma\gamma}$ and $\mu_{gg}^{ZZ}$
in the Type-I 2HDMs.
}
\label{fig1}
\end{figure}
%--------------------------------------------------------------------------

In this work we consider the Type-I fermiophobic 2HDMw$U(1)_H$.
There are many theoretical and experimental constraints on the model 
parameters.
The Higgs potential must satisfy constraints from conditions on 
perturbativity, unitarity and vacuum stability. We also take into account 
constraints from electroweak precision observables, exotic top decay,
$b\to s\gamma$, heavy Higgs boson search at the LHC, and SM-like Higgs 
boson search at the LHC. Finally, if the SM-like Higgs boson can decay
into non-SM particles, search for invisible Higgs decay at the LHC 
also strongly constrain the parameters in the 2HDMw$U(1)_H$~\cite{2hdmu1}.
If both Higgs doublets develop VEVs, there is a tree-level mixing
between $Z$ and $Z_H$. If one of them does not develop a VEV, there is 
no mixing between $Z$ and $Z_H$ at the tree level.
However in general, the mixing can be generated at the one-loop level, and
the mixing angle $\xi$ is strongly constrained by experiments:
%. The mixing angle $\xi$
%should be 
$\sin\xi \lesssim O(10^{-2}) \sim O(10^{-3})$.  
Figure~\ref{fig1} shows the signal strengths  $\mu_{gg}^{\gamma\gamma}$ 
and $\mu_{gg}^{ZZ}$ in the Type-I 2HDMs.
The red and blue points satisfy all constraints in the 2HDM with $Z_2$ symmetry
(2HDMw$Z_2$) and 2HDMw$U(1)_H$, respectively. Both models are consistent with
CMS data in the $1\sigma$ level,  but with ATLAS data in the $2\sigma$ level.
In the region $\mu_{gg}^{ZZ} \lesssim 0.4$, 
the 2HDMw$U(1)_H$ could be distinguished from the 2HDMw$Z_2$.
If the Higgs boson properties are found to be close to the SM prediction,
it would be essential to discover the extra scalar boson and new gauge boson
to distinguish the 2HDMw$U(1)_H$ from the 2HDMw$Z_2$ as well as from the SM.

\section{Inert doublet Model (IDM)}
There are many evidences for the existence of nonbaryonic cold dark matter (CDM) 
in our universe.  Among many models for CDM, a weakly interacting 
massive particle (WIMP) is  an interesting scenario. In the 2HDMw$Z_2$,
the dark matter candidate can be one of extra scalars when one of Higgs 
doublets does not develop a VEV and an exact $Z_2$ symmetry is imposed.
This modes is called as the inert doublet model (IDMw$Z_2$)~\cite{inert}.
Under the $Z_2$ symmetry, SM particles including the ordinary Higgs doublet 
responsible for electroweak symmetry breaking  are $Z_2$-even. 
%And the new Higgs doublet with $H^+$, $H$, and $A$ is $Z_2$-odd. 
%One of Higgs doublet is responsible for the electroweak symmetry
%breaking 
And the other Higgs doublet has $Z_2$-odd particles, $H^+$, $H$, and $A$. 
In this work, we assume the neutral $Z_2$-odd  scalar $H$ is dark matter.
In the Higgs potential the dim-2 terms ($H_1^\dagger H_2 + h.c.$) do not appear
because of the exact $Z_2$ symmetry.

As in the 2HDMw$U(1)_H$, the $Z_2$ symmetry can be replaced 
by local $U(1)_H$ gauge symmetry~\cite{idmu1}.
In this case, we have to add a SM-singlet scalar $\Phi$. Without $\Phi$, the $U(1)_H$
symmetry would be unbroken, resulting in  massless  $Z_H$ boson.
The remnant symmetry of $U(1)_H$ is the origin of the exact $Z_2$ symmetry 
stabilizing dark matter $H$.
The dim-2 terms $H_1^\dagger H_2 + h.c$ are forbidden by $U(1)_H$ symmetry. 
In addition,  the  $\lambda_5$ terms $(H_1^\dagger H_2)^2 + h.c.$ in the usual 
inert doublet models are also forbidden by the $U(1)_H$ symmetry. 
Without $\lambda_5$ terms, the neutral boson $H$ and pseudoscalar boson $A$ are 
degenerate. Then the cross section for $H N \to A N$, where $N$ is a nucleon, 
through the $Z$ boson exchange could be large so that 
search for direct detection of dark matter in the XENON100 or LUX experiments
would  exclude this model immediately.
The $\lambda_5$ terms can be generated effectively  by a higher-dimensional 
operator (see Ref.~\cite{idmu1} for more detail). 
And after $U(1)_H$ symmetry breaking, small $\lambda_5$ terms are induced and 
generate mass difference between $H$ and $A$ so that the $H N \to A N$ process is
kinematically forbidden.

%--------------------------------------------------------------------------
\begin{figure}[t]
\begin{minipage}{0.5\linewidth}
\centerline{\includegraphics[width=0.9\linewidth]{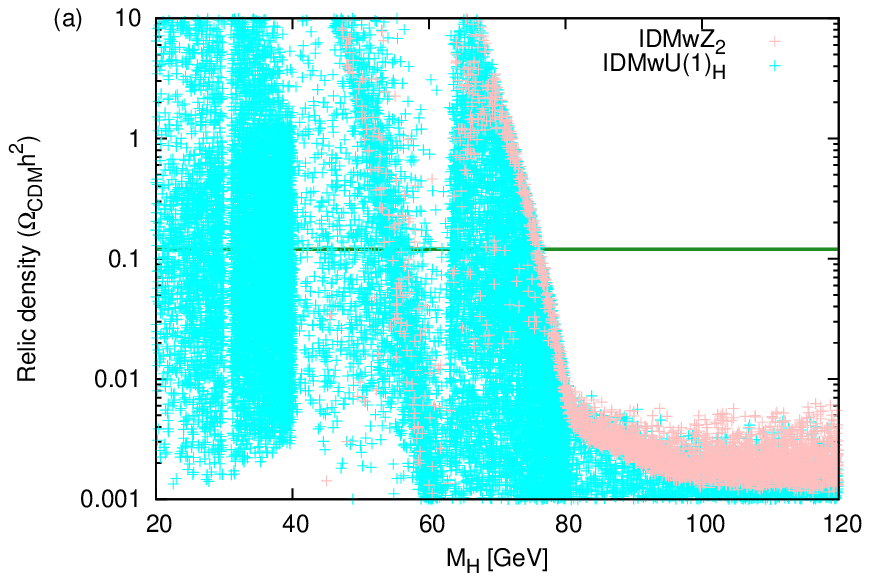}}
\end{minipage}
\hfill
\begin{minipage}{0.5\linewidth}
\centerline{\includegraphics[width=0.9\linewidth]{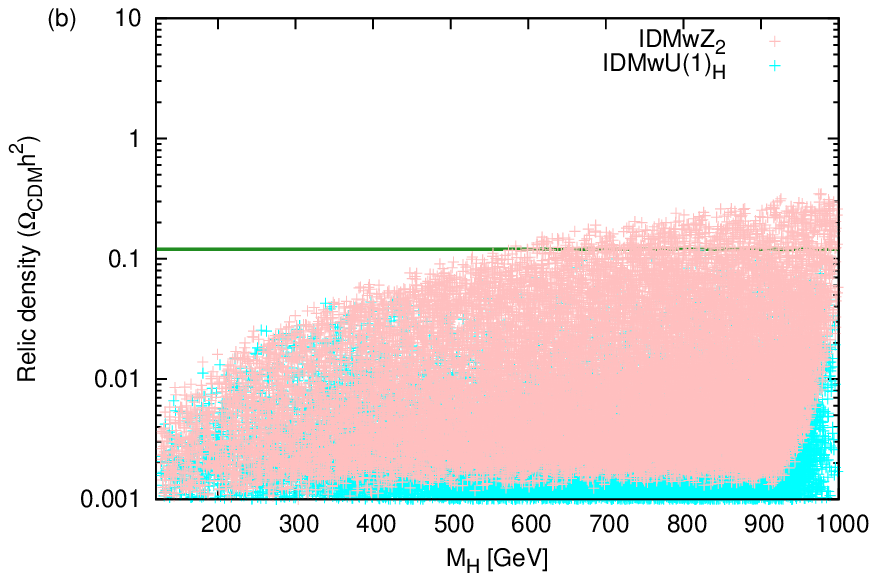}}
\end{minipage}
\hfill
\caption[]{
Dark matter mass and relic density (a) in the light $H$ scenario and
(b) in the heavy $H$ scenario.
}
\label{fig2}
\end{figure}
%--------------------------------------------------------------------------

Figure~\ref{fig2} shows the dark matter mass $m_H$ and relic density
(a) in the light $H$ scenario and (b) in the heavy $H$ scenario.
The pink points are in the IDMw$Z_2$ while the cyan points are
in the IDMw$U(1)_H$, respectively. 
All the points satisfy constraints from LUX experiments
and the horizontal line is the current experimental value for the relic density.
As shown in Fig.~\ref{fig2} (a),  a larger parameter space is allowed 
in the IDMw$U(1)_H$ in the light $H$ scenario.
In particular, a light CDM scenario ($m_H \lesssim 40$ 
%GeV is possible, but in the IDMw$Z_2$ a light CDM is ruled out.
%However, in the IDMw$U(1)_H$, the light CDM scenario with $m_H \lesssim 40$ 
GeV) is still possible, because the $HH\to Z_H Z_H$ and $HH\to Z Z_H$
processes mainly contribute to the relic density. This is a new aspect of 
IDMw$U(1)_H$, which was not possible in ordinary IDMw$Z_2$.
Near the SM-Higgs resonant region, $m_H \sim 60-80$ GeV,
the co-annihilation of $H A$ or $H H^+$ and the pair annihilation of
$AA$ and $H^+ H^-$ also contribute to the relic density.
In the heavy $H$ scenario, both models are not so different 
from each other because the annihilation cross sections for
$HH\to WW$ and $HH\to ZZ$ are dominant ones.  In most cases
the predicted relic density in the IDMw$U(1)_H$ is slightly less than
that in the IDMw$Z_2$ because of new channels $HH\to Z_H Z_H$
and $HH\to Z Z_H$.

%--------------------------------------------------------------------------
\begin{figure}[t]
\begin{minipage}{0.5\linewidth}
\centerline{\includegraphics[width=0.9\linewidth]{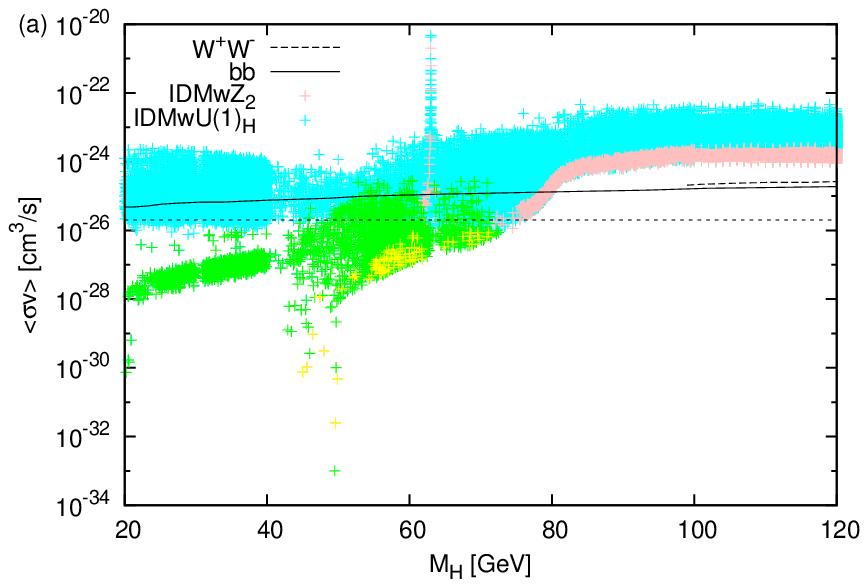}}
\end{minipage}
\hfill
\begin{minipage}{0.5\linewidth}
\centerline{\includegraphics[width=0.9\linewidth]{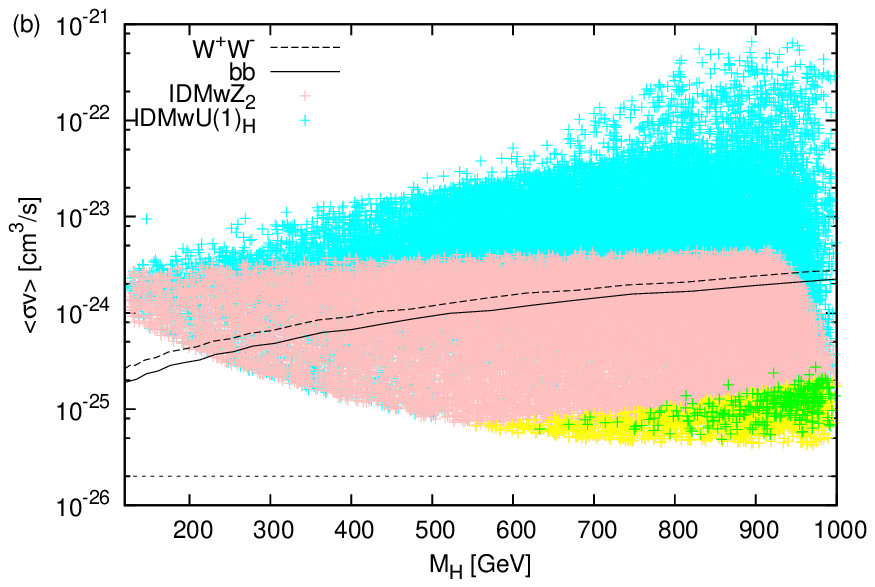}}
\end{minipage}
\hfill
\caption[]{
$M_H$ and $\langle \sigma v \rangle$ (a) in the light $H$ scenario and
(b) in the heavy $H$ scenario.
}
\label{fig3}
\end{figure}
%--------------------------------------------------------------------------

Figure~\ref{fig3} shows $m_H$ and the velocity-averaged annihilation 
cross section, $\langle \sigma v \rangle$ at present
(a) in the light $H$ scenario and $(b)$ in the heavy $H$ scenario.
The pink points are in the IDMw$Z_2$ while the cyan points are
in the IDMw$U(1)_H$, respectively. 
The lower horizontal line comes from constraint on the $S$-wave dark matter
annihilation from the relic density observation, 
while the upper two curves denote constraints on $\langle \sigma v \rangle$
from Fermi-LAT's analysis of 15 dwarf spheroidal galaxies.
Two curves correspond to the $HH\to WW$ and $HH\to b\bar{b}$ dominant cases,
respectively. The IDMw$U(1)_H$ is more complicated than the naive assumption
so that we cannot apply the above constraints to the our models.
We calculate the quantity $\Phi_\textrm{PP}$, which is the part of
the gamma ray flux from dark matter annihilation~\cite{gamma}.
The 95\% upper bound is given by 
$9.3 \times 10^{-30}$ cm$^3$s$^{-1}$GeV$^{-2}$~\cite{gamma}.
The yellow and green points in Fig.~\ref{fig3} satisfy this upper bound
in the IDMw$Z_2$ and in the IDMw$U(1)_H$, respectively.
In the light $H$ scenario, the region $m_H \lesssim 40$ GeV is allowed
only in the IDMw$U(1)_H$. Actually in this region, only $HH\to Z_H Z_H$
contribute to $\langle \sigma v \rangle$ and $m_H \sim m_{Z_H}$.
In the heavy $H$ scenario, the allowed region appears at $m_H \gtrsim$ 500
GeV. For more detailed analysis we refer the reader to Ref.~\cite{idmu1}.

\section{Conclusions}
Two Higgs double models are interesting extensions of the SM and  appear in many 
high-scale theories. 
%, and may be effective 
%models of high-energy theories and useful to test the underlying theory.
We showed that the Higgs mediated FCNC problem in 2HDM could be  resolved 
by considering gauged $U(1)_H$ symmetry instead of discrete $Z_2$ symmetry.
After $U(1)_H$ symmetry breaking, the remnant symmetry would be the origin of the 
usual $Z_2$ symmetry.
This $U(1)_H$ extension can also be applied to the IDM with exact $Z_2$ symmetry.  
%where an exact $Z_2$ symmetry is imposed.
The stability of dark matter in IDM with $U(1)_H$
symmetry is guaranteed by the remnant $Z_2$ symmetry of $U(1)_H$.
We showed that in the type-I case, a light CDM scenario,
which is ruled out in the typical IDM, is possible in the IDM
with $U(1)_H$ symmetry.

\section*{Acknowledgments}

This work was supported by Basic Science Research Program through 
NRF (2011-0022996).

\end{document}